# Development and Assessment of a Miniaturized Thermocouple for Precise Temperature Measurement in Biological Tissues and Cells

Onnop Srivannavit [1], Rakesh Joshi [1], Weibin Zhu [1], Bin Gong [3], Stuart C. Sealfon [1], Theodorian Borca-Tasciuc [4], Angelo Gaitas [1, 2]

[1] The Estelle and Daniel Maggin Department of Neurology, Icahn School of Medicine at Mount Sinai, New York, NY 10029, USA.

[2] BioMedical Engineering & Imaging Institute, Leon and Norma Hess Center for Science and Medicine, Icahn School of Medicine at Mount Sinai, New York, NY 10029, USA.

[3] The University of Texas Medical Branch at Galveston, Galveston, TX, USA

[4] Mechanical, Aerospace & Nuclear Engineering, School of Engineering, Rensselaer Polytechnic Institute, Jonsson Engineering Center, Troy, NY, USA

**Abstract:**

This study presents a novel thermocouple instrument designed for precise temperature monitoring within biological tissues and cells, addressing a significant gap in biological research. Constructed on a Silicon-On-Insulator (SOI) substrate, the instrument employs doped silicon and chromium/gold junctions, achieving a Seebeck coefficient of up to 447 µV/K, rapid response times, high temperature accuracy, and the necessary durability for tissue measurements. The cleanroom fabrication process yields a device featuring a triangular sensing tip. Using Finite Element Analysis (FEA) with COMSOL Multiphysics, the research delves into the device's thermal time constant within tissue environments. The device's efficacy in biological settings was validated by measuring temperatures inside ex-vivo tissue samples. Our findings, bolstered by FEA COMSOL simulations, confirm the device's robustness and applicability in biological studies. This advancement in thermocouple microneedle technology provides biologists with an instrument for accurately tracking temperature fluctuations in tissues.

## 1. Introduction

Understanding temperature gradients is crucial in the field of biology, particularly in comprehending biological functions and various disease states[1-19]. Temperature fluctuations are integral to processes such as metabolism, cell division, and other cellular activities.[20,21] Cancer cells generate heat [22] and temperature is raised in response to infections, triggering innate and adaptive immune responses[23]. Brown adipose tissue is known to produce heat in response to cold conditions[24,25]. Furthermore, neuroscience research has highlighted that even slight brain temperature variations of 1-2 °C can significantly impact memory and behavioral patterns[26].

Techniques involving molecular reporters have been employed to gauge temperature gradients within cells, revealing differences of approximately 1°C to 2.9 °C between cell nuclei and cytoplasm, depending on the measurement method utilized [1-4]. Remarkably, temperature differentials of up to 10°C originating from mitochondria have been reported using temperature-sensitive fluorescent probes[5-7].

A key tool for measuring temperatures is the thermocouple, which consists of two different electrically conductive materials, typically metals joined at a junction while their free ends are connected to a voltage measuring apparatus[27]. Temperature changes at this junction generate an electromotive force proportional to the temperature difference between the junction and the free end (named the cold junction) for each metal, a phenomenon known as the Seebeck effect. Typically, thermocouples are made from pairs of materials with positive and negative signs of the Seebeck coefficient in order to increase the measured voltage. The distinct advantage of a thermocouple lies in its sensitivity to temperature changes solely at the thermocouple junction, rendering it unaffected by temperature fluctuations in other parts of the device[27] (assuming known cold junction temperature). Thermocouples are also characterized by ease of calibration, insensitivity to fluid viscosity and chemical processes, non-toxicity to cells, and rapid response times.

However, most micromachined thermocouples are not designed for use in mammalian tissue measurements and often lack the necessary electrical insulation for operation in such environments [28-34]. Recent advancements have seen the development of standard metal junction thermocouples for biological applications, including a tungsten-platinum thermocouple device[35] and a metal junction thermocouple used in measuring temperatures in large single cells such as in neurons of Aplysia californica[26,36].

In contrast to standard metal junction thermocouples, which exhibit relatively modest Seebeck coefficients (<100µV/K combined Seebecks), doped Si stands out for its capacity to reach significantly higher values[31,32,37-41]. Notably, intrinsic Si, in its undoped state, can achieve Seebeck coefficients around -1000 µV/K. The process of doping Si, wherein impurities are introduced to alter its properties, has a marked impact on its Seebeck coefficient. Light doping can result in coefficients within the range of several hundred µV/K. However, as the doping level increases, there is a notable reduction in the Seebeck coefficient, potentially dropping to tens of µV/K or lower in heavily doped Si.

This paper presents the fabrication of a thermocouple device, utilizing doped Si and Cr/Au junctions on a SOI wafer. This novel thermocouple device distinguishes itself from current state-of-the-art technologies through several key advantages, particularly beneficial for biological applications. First, the miniaturized design of our device, enabled by micromachining methods (micro-electromechanical systems - MEMS) and the use of doped Si and Cr/Au junctions, allows for precise temperature measurements within the constrained and sensitive environments of tissues and large cells. The design is also will allow further miniaturization in the future. The needle-like design enhances its application by enabling effective penetration into tissues, with its structural integrity allowing for measurements at depths of 1.5 mm and beyond without compromising the device's performance. The higher combined Seebeck coefficient of the junction of up to 447 µV/K achieved in our device translates to enhanced sensitivity in detecting temperature variations, a critical factor in biological research where even minor temperature changes can have significant implications. Moreover, the rapid response time of our thermocouple facilitates real-time monitoring of dynamic biological processes, a capability often limited by the slower response times of conventional thermocouples. Compared to existing solutions, which may lack the necessary electrical insulation for operation in biological environments, our device is electrically insulated. This study establishes a solid groundwork for developing thermocouples capable of performing precise temperature measurements within tissues and large cells.

## 2. Device fabrication

The device (***Fig.1a***) includes a triangular shaped junction made of Cr/Au (40nm/160nm) and doped Si near the tip with an area of 54 µm$^2$ (***Fig. 1b***) and a second reference junction on the chip handle. The cantilevers have widths of 40 µm, thicknesses of 10 µm, and lengths of 500 µm, 1000 µm, and 1500 µm. The device fabrication steps are shown in ***Fig. 2***. For the thermocouple junction formation, we followed the fabrications steps described previously [38,39]. First, the SOI wafers were cleaned using a Nanostrip solution for 10 minutes, this was followed by BF2 ion implantation. We then cleaned the wafers with Complementary Metal-Oxide-Semiconductor (CMOS) organic and ionic cleans, followed by an annealing session at 850°C for 30 minutes ***(Fig. 2a)***. After the annealing step, the sheet resistivity of the wafers was measured using a four-point probe instrument. This was then followed by Plasma Enhanced Chemical Vapor Deposition (PECVD) for oxide deposition with average thickness of 100nm on the front side of the wafers (***Fig. 2b***). Then windows for the junction were formed by photolithography and dry oxide etching (***Fig. 2c***). Photolithography of metal patterns is followed (***Fig. 2d***) by the metal deposition of 40 nm thick Cr and 160 nm thick Au and lift off processes (***Fig.2e***). Front-side cantilevers were formed by photolithography and then Si deep reactive ion etching (DRIE) (***Fig. 2f***). A 30 nm thick insulation layer of Al$_2$O$_3$ was deposited. Then, pad openings in the Al$_2$O$_3$ layer for electrical connections were created by photolithography and BHF oxide etching. Backside cantilevers were formed using photolithography and Si DRIE to remove the handle layer and dry oxide etching to remove Buried Oxide layer (BOX) ***(Fig. 2g)***.

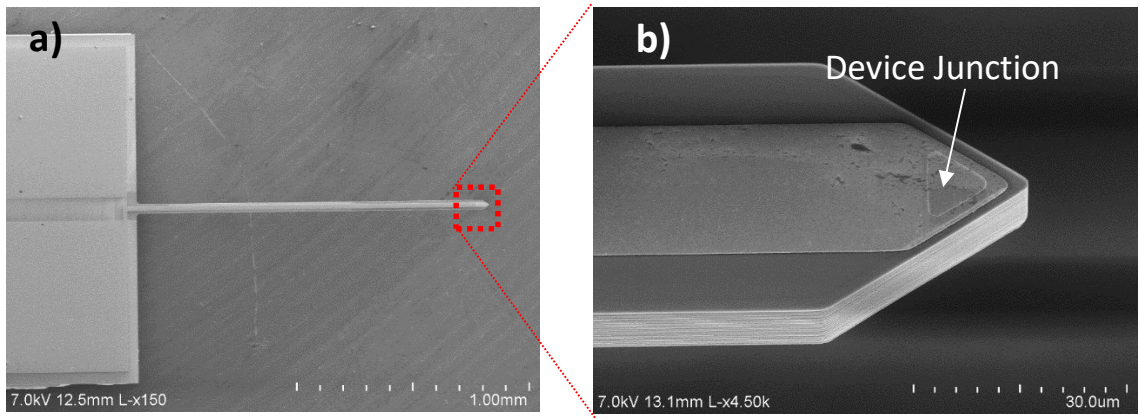

***Fig. 1.*** (a) Scanning Electron Microscopy (SEM) image of 1500 um long device. (b) Close-up SEM image of the triangular thermocouple junction at the tip of the cantilever.

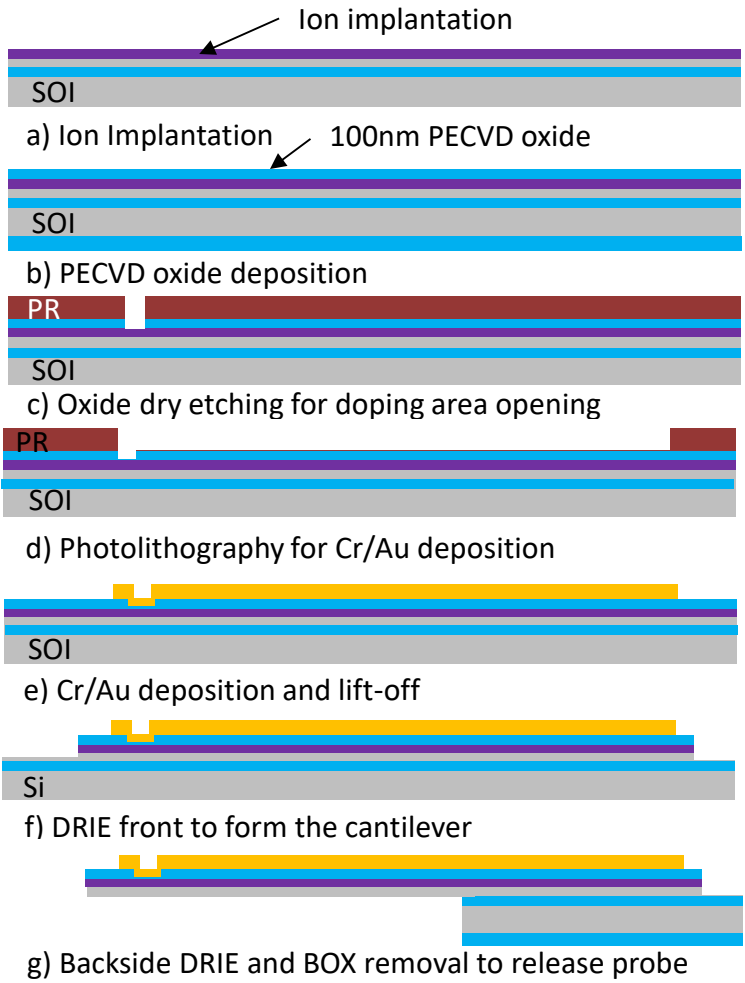

Fig. 2. The microfabrication process flow of the micromachined thermocouple.

## 3. Device characterization

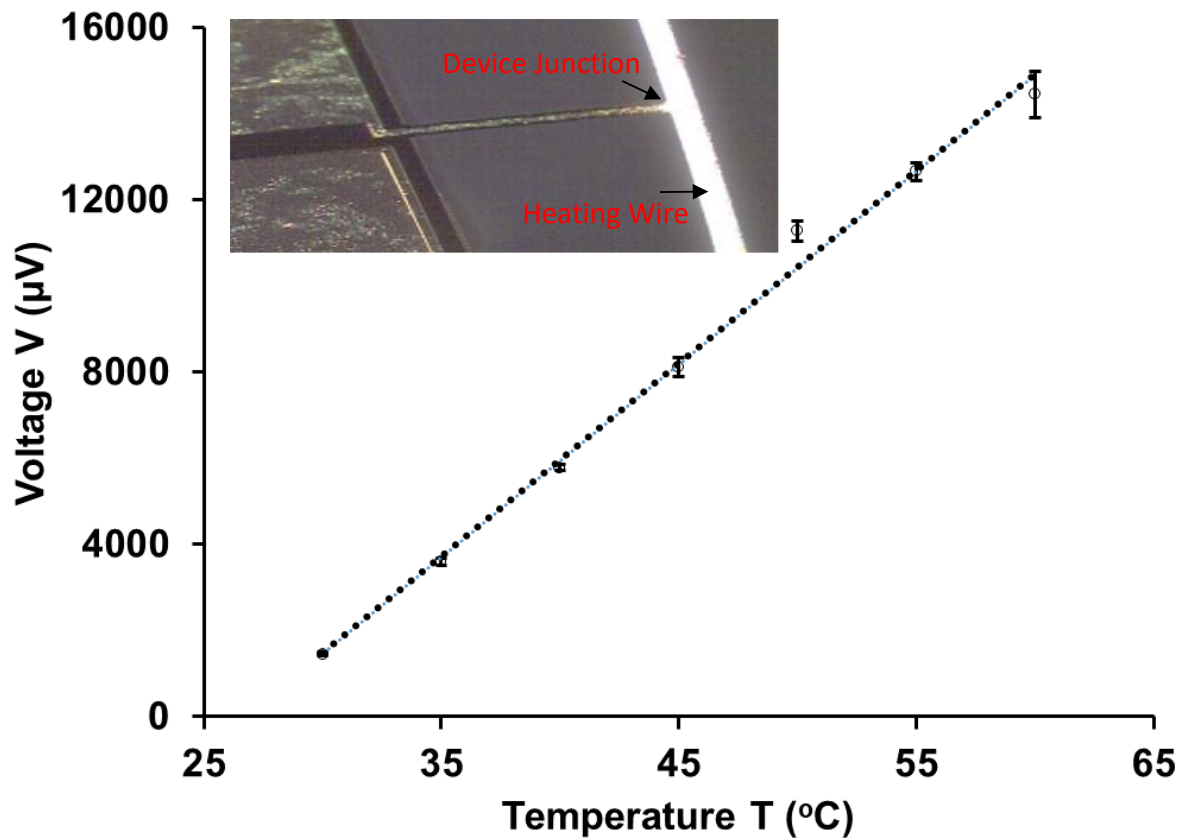

*Fig. 3* Measurement of a thermocouple device voltage (V) vs temperature (T) of a pre-calibrated thin wire (for each V measurement: N=10) used to derive Seebeck coefficient 447 µV/°C, (*in-set*) optical image of the device in contact with calibration wire in air.

In order to measure the Seebeck coefficient, a titanium wire of 50µm diameter was utilized, which was heated using a Keithley 2400 sourcemeter (Tektronix, United States). Calibration of this system was conducted employing an Omega thermocouple (HH506RA, Omega Engineering Inc, USA), having a diameter of 30 µm. Subsequently, a 34401A multimeter (Keysight, USA) was incorporated into the setup to record voltage changes in the probe thermocouple. The experimental procedure involved bringing the thermocouple into proximity with the pre-calibrated wire, as depicted in *Fig. 3 (in-set)*. A gradual heating process was initiated for the wire by increasing the current using the sourcemeter, during which both the voltage of the thermocouple and temperature of the wire were concurrently recorded. The thermocouples exhibited a Seebeck coefficient ranging from 141.4±11.9 µV/°C and 447.3±16.6 µV/°C (*Fig. 3*), aligning with previously reported range of values[38,39]. The temperature resolution of our measurement was ascertained by dividing the multimeter's resolution limit of 0.1 µV by the Seebeck coefficient, resulting in a temperature resolution between 0.00024°C and 0.0007°C. The observed variations in the Seebeck coefficient from one device to another are likely due to heterogeneity in silicon doping, non-uniformity in manufacturing processes, and the influence of surface effects and oxides at the junction. The Seebeck coefficient of commercial thermocouples is significantly lower. For instance, the commonly used K-type thermocouples exhibit a Seebeck coefficient of approximately 40 µV/°C, while the E-type thermocouples, known for having one of the highest Seebeck coefficients commercially available, display a coefficient around 68 µV/°C.

The thermal time constant of the thermocouple device within tissue was determined through time-dependent numerical simulations using COMSOL Multiphysics 5.5 FEA. The model involved a solid tissue, where the temperature of the tissue was lower and matched that of the surrounding environment at 293 K. The dimensions of the tissue sample were specified as 2 mm by 2 mm. A convective heat flux boundary condition was applied at the interface between the tissue and the environment, assuming a heat transfer coefficient of 10 W/(m^2·K), akin to that of air. The thermocouple device was initially set to a temperature of 313 K. Material properties for silicon (Si), gold (Au), silicon dioxide (SiO2), and tissue are detailed in Table 1. The analysis was conducted in two dimensions. Examples of the cantilever's temperature profile at time values t=0, $10^{-5}$, and $10^{-4}$ s are depicted in *Fig. 4a, b, and c* showing a uniform distribution of temperature across the cross section of the cantilever. In *Fig. 4d* a magnification of the Au/Si junction area at t=$10^{-5}$ s is shown. The thermal time constant was estimated to be approximately 57 µs, calculated from the temperature-time graph at the Au-Si junction interface (*Fig. 4e*), by identifying the point at which the temperature difference decreased to 1/e of its initial value.

**Table 1.**

|  | Si | Au | SiO2 | Tissue |
|---|---|---|---|---|
| Heat capacity at constant pressure, Cp (J/(kg·K)) | 700 | 129 | 730 | 3540 |
| Density, ρ (kg/m³) | 2329 | 19300 | 2200 | 1079 |
| Thermal conductivity, k (W/(m·K)) | 130 | 317 | 1.4 | 0.52 |

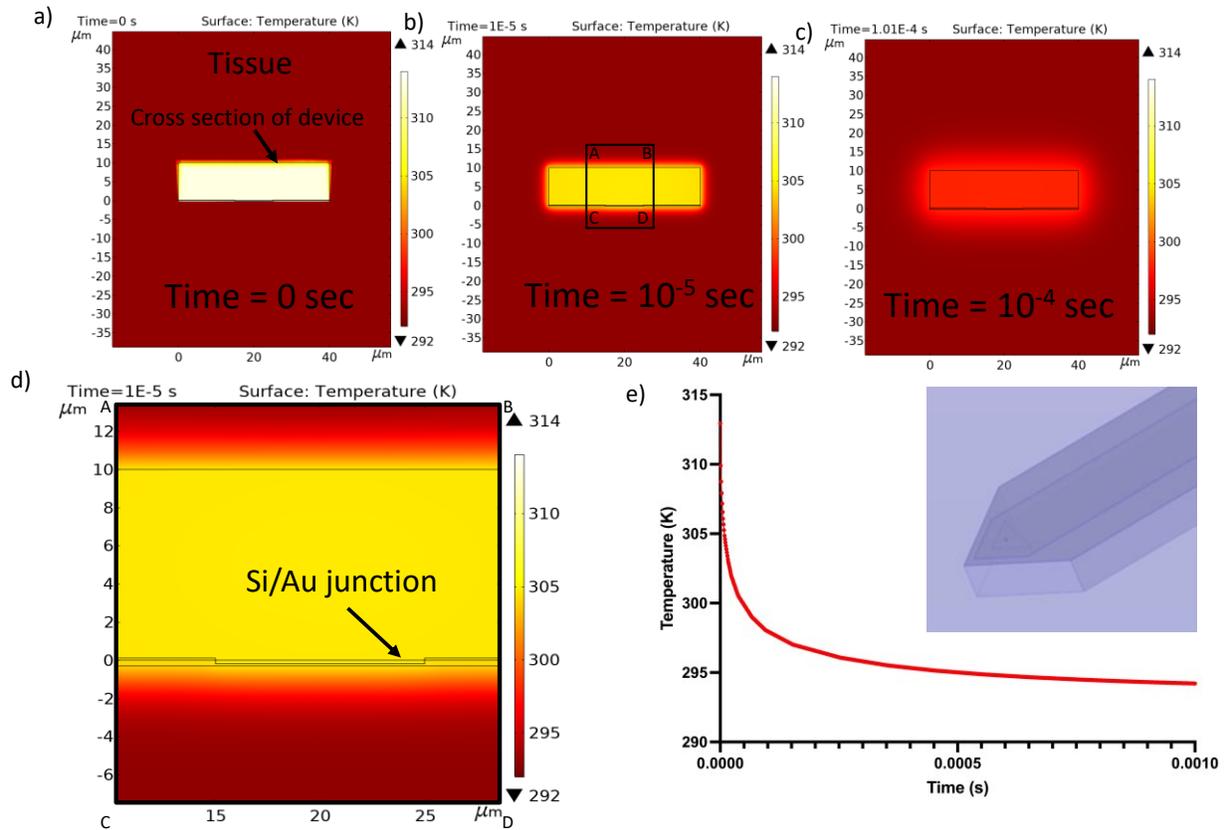

*Fig. 4.* FEA simulations of the temperature distribution across the device's cross-section. (a) Initial conditions with device temperature at 313 K and adjacent tissue at 293 K. (b) Temperature at time $10^{-5}$ s, shows that temperature distribution remains uniform. (c) Temperature profile at time $10^{-4}$ s. (d) Zoomed-in view of the temperature distribution at $10^{-5}$ s. (e) Graph of temperature versus time at the Au-Si junction interface, highlighted by an arrow in (d). The thermal time constant was determined by calculating the time required for the temperature to decrease from its initial value to $1/e$ of that value.

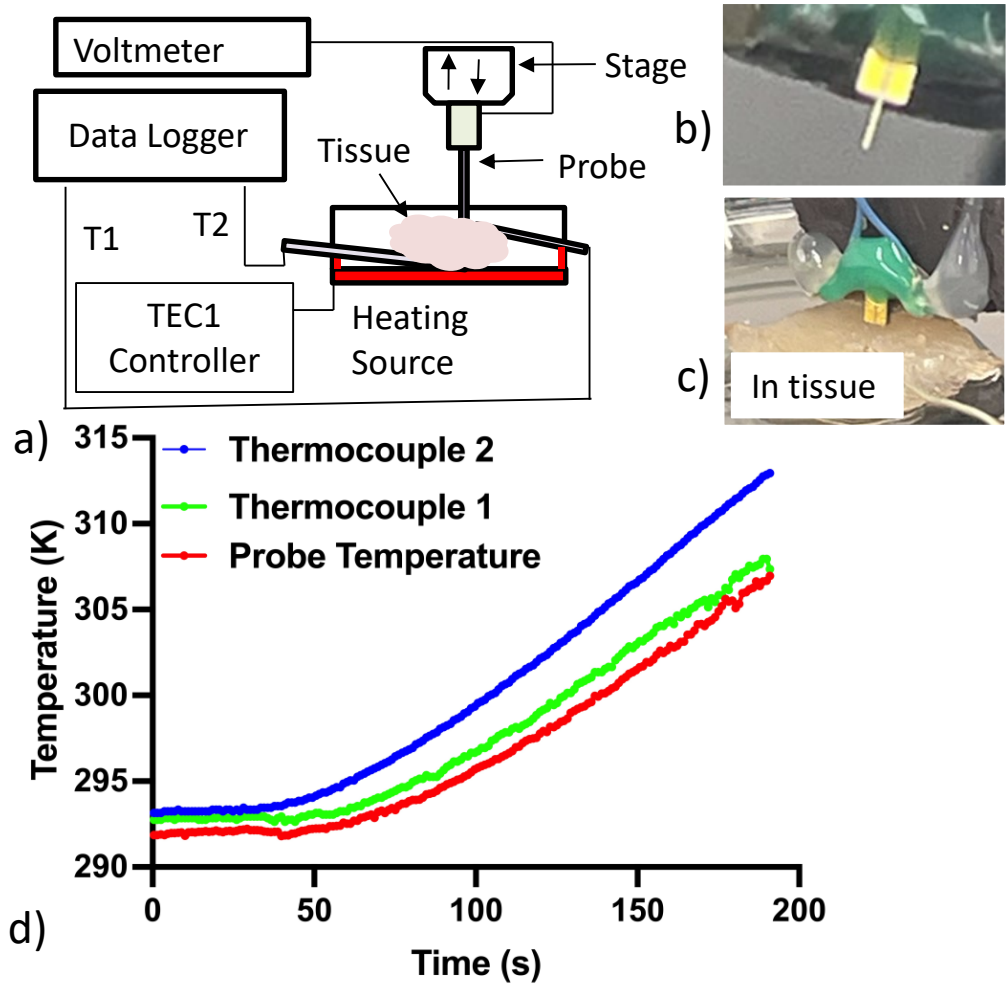

*Fig. 5.* Temperature measurements on ex-vivo tissue. (a) Experimental Setup. T1 and T2 are commercial Omega thermocouples. T1was placed on top and inside the tissue via a small incision and T2 was placed under the tissue and in direct contact with the heater. Our probe was inserted in the tissue as shown using a micromanipulator. (b) Probe advancing towards the tissue. (c) Probe is inserted into the tissue. (d) Temporal temperature rise recorded by the probe and two commercial thermocouples (thermocouple 1 - T1 in and thermocouple 2 - T2) positioned at distinct tissue sites.

To assess the performance of our device within a biological tissue setting, we performed temperature measurements on chicken breast tissue. The TEC-CH1 controller (Nanosurf, Switzerland) served as the heating element to deliver heat to the tissue. Changes in voltage of our device were captured using the previously mentioned voltmeter, as depicted in the experimental setup (*Fig. 5a*). Following the insertion of the thermal probe into the tissue (*Fig. 5b*), we increased the heater's temperature to monitor the tissue's internal temperature changes. These changes were recorded using our probe alongside two Omega thermocouples, described previously. The Omega thermocouples were placed at the bottom of the dish under the tissue and within the tissue itself, as illustrated in *Fig. 5a*. With the probe attached to a holder and inserted into the tissue, it was possible to simultaneously track temperature readings from both the probe and thermocouples

(*Fig. 5bc*). The repeated insertion of the probe into the tissue verified its durability and structural integrity (*Fig. 5c*). Analysis of the graph in *Fig. 5d* reveals that the three thermocouple devices displayed similar behavior in terms of temperature increase, demonstrating that our device is capable of accurately recording temperature changes within tissues. This indicates not only the device's effectiveness in measuring temperature but also its unique ability, unlike existing micro thermocouples, to successfully penetrate tissue.

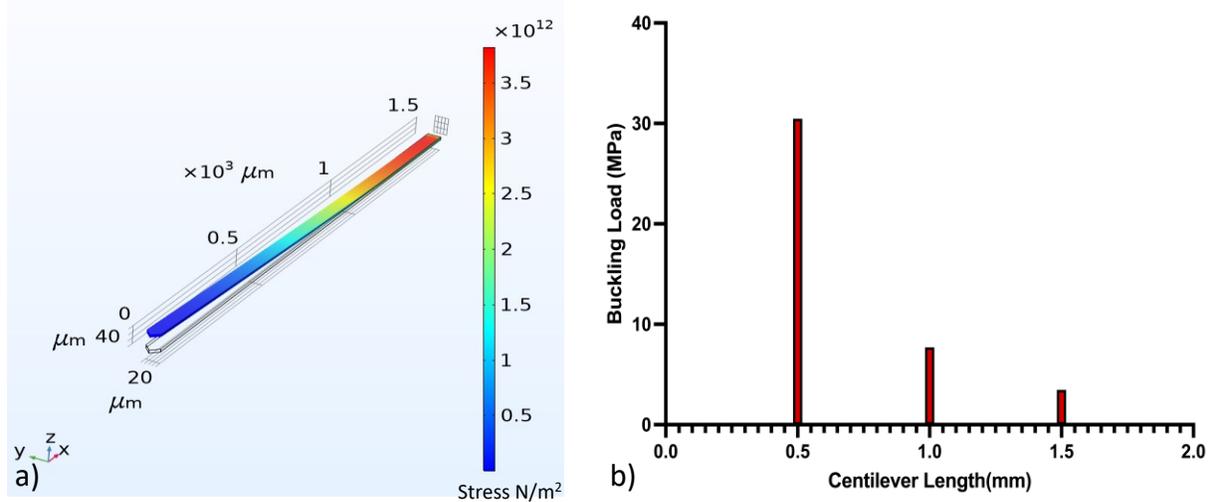

*Fig. 6* (a) An FEA model utilizing the solid mechanics module was used to simulate the stress distribution. The graphical output employs color contours to depict the Von Mises stress magnitude throughout the structure. Regions experiencing elevated stress levels are highlighted in red, signifying higher intensity, whereas zones of diminished stress are conveyed through cooler (blue) tones. (b) Buckling load vs. cantilever length.

In our study, experimental observations confirmed the structural integrity of the device. To further validate these findings, we employed COMSOL Multiphysics 5.5, using its solid mechanics capabilities to conduct an analysis on a three-dimensional representation of the device, as depicted in *Fig. 6a*. The device is a cantilever with the free end at the tip and fixed end at the base. We subjected the model to a load along the x-axis to ascertain its linear buckling threshold [42,43]. This analysis enabled us to calculate the critical buckling load, which is the lowest pressure that induces buckling in the device. Our findings (*Fig. 6b*) revealed that for a device length of 1.5 mm, the critical buckling load was 3.46 MPa. For a device measuring 1 mm in length, this value increased to 7.71 MPa, and for a length of 0.5 mm, the critical buckling load further increased to 30.47 MPa. These findings are particularly significant when compared against the typical resistance of human skin, which is approximately 3.18 MPa [43-45]. This comparison suggests that all our device variations possess the mechanical capability to penetrate the skin, a conclusion that is also supported by our experimental data. This analysis focuses solely on the aspect of insertion pressure. However, for a comprehensive assessment of the device's stability, it is essential to take into account additional factors such as bending forces and various other potential stresses scenarios during insertion into the skin, which could result in mechanical failure [43,46,47].

## 4. Conclusions

In this study, we have introduced a novel thermocouple microneedle device designed on an SOI wafer that significantly advances the capability for precise temperature measurements within biological tissues and large cells. This device, featuring a unique combination of doped Si and Cr/Au junction, demonstrates a high Seebeck coefficient, rapid response times, and adequate structural integrity, making it ideal for use in biological environments. The efficacy of this device has been validated through comprehensive characterizations and FEA, which also provided insights into its thermal behavior in tissue contexts. Experimental application in ex-vivo tissues further confirmed its accuracy and suitability for biological research. While our mechanical analysis primarily focused on insertion pressure, recognizing the importance of other stress factors is crucial for a complete understanding of the device's stability in practical use.

## Acknowledgements

This work was supported by the National Science Foundation under award number 226930, the National Institute of General Medical Sciences under award number R44GM146477; the content is solely the responsibility of the authors and does not necessarily represent the official views of the National Institutes of Health or the National Science Foundation. The authors extend their sincere appreciation to Prof. Takahito Ono from the Department of Mechanical Systems Engineering at Tohoku University, Sendai, Japan, for his guidance in thermocouple microfabrication.

**Keywords**: Thermocouple Device, Biological Temperature Measurement, Seebeck Coefficient, microfabrication, Structural Integrity.